\documentclass{aastex61}

\newcommand\aastex{AAS\TeX}

\usepackage{soul}

\received{}
\revised{}
\accepted{}
\submitjournal{ApJ}

\shorttitle{\aastex\ turbulence in jets collision}
\shortauthors{Pucci et al.}

\begin{document}

\title{Generation of turbulence in colliding reconnection jets}

\correspondingauthor{Francesco Pucci}
\email{francesco.pucci@kuleuven.be}

\author{Francesco Pucci}
\affiliation{Center for Mathematical Plasma Astrophysics, Department Wiskunde, KU Leuven\\
200B Celetjinenlaan\\
Leuven, B-3001, Belgium}

\author{William H. Matthaeus}
\affiliation{Department of Physics and Astronomy, University of Delaware\\
217 Sharp Lab\\
Newark, DE 19716, USA}

\author{A. Chasapis}
\affiliation{Department of Physics and Astronomy, University of Delaware\\
217 Sharp Lab\\
Newark, DE 19716, USA}

\author{Sergio Servidio}
\affiliation{Dipartimento di Fisica, Universit\`a della Calabria\\
Via P. Bucci, Cubo 31C\\
Arcavacata di Rende, I-87036, Italy}

\author{L. Sorriso-Valvo}
\affil{Nanotec-CNR, Sede di Cosenza\\ 
Via P. Bucci, Cubo 31C\\ 
Arcavacata di Rende, 87036, Italy}

\author{V. Olshevsky}
\affil{Center for Mathematical Plasma Astrophysics\\ 
Department Wiskunde, KU Leuven, 200B Celestijnenlaan\\
Leuven, 3001, Belgium}

\author{D. L. Newman}
\affil{University of Colorado\\
Boulder, CO 80309, USA}

\author{M. V. Goldman}
\affil{University of Colorado\\
Boulder, CO 80309, USA}

\author{Giovanni Lapenta}
\affiliation{Center for Mathematical Plasma Astrophysics, Department Wiskunde, KU Leuven\\
200B Celetjinenlaan\\
Leuven, B-3001, Belgium}

\begin{abstract}
The collision of magnetic reconnection jets is studied by means of a three 
dimensional numerical simulation at kinetic scale, in the presence of a strong 
guide field. We show that turbulence develops due to the jets collision 
producing several current sheets in reconnection outflows, aligned with the 
guide field direction. The turbulence is mainly two-dimensional, with stronger 
gradients in the plane perpendicular to the guide field and a low wave-like 
activity in the parallel direction. First, we provide a numerical method to 
isolate the central turbulent region. Second, we analyze spatial second-order 
structure function and prove that turbulence is confined in this
region. Finally, we compute local magnetic and electric frequency 
spectra, finding a trend in the sub-ion range that differs from typical cases 
for which the Taylor hypothesis is valid, as well as wave activity in the range between 
ion and electron cyclotron frequencies. 
Our results are relevant to understand observations of reconnection jets 
collisions in space plasmas.

\end{abstract}


\section{Introduction} \label{sec:intro}

Magnetic reconnection is a fundamental phenomenon in 
astrophysical plasmas. It consists in
the recombination of magnetic field topology due 
to the violation of the frozen-in law for magnetic
field of ideal magnetohydrodinamics (MHD).  
Due to this recombination, plasma is ejected 
in the form of jets and magnetic energy is 
converted into kinetic energy. The presence
of magnetic reconnection has been ascertained in the 
solar corona \citep{sui2004evidence,shibata1998evidence}, in the solar wind 
\citep{gosling2007observations,phan2006magnetic,retino2007situ} 
and in the Earth's magnetosphere \citep{burch2016electron},
and it is thought to be responsible for particle
acceleration and plasma heating in these environments \citep{drake2006electron}. 
A close link exists between the presence of magnetic 
reconnection and the phenomenology of plasma turbulence \citep{matthaeus2011needs}. 
In plasmas, similarly to fluid dynamics, magnetic and velocity fluctuation 
energy cascade 
from large to small scales, due to non-linear
interactions, following the turbulent phenomenology. 
This non-linear cascade produces smaller and smaller
scale magnetic shears, that can eventually undergo the process of 
magnetic reconnection.
Magnetic reconnection, indeed,
can be viewed as an active ingredient of plasma turbulence 
\citep{matthaeus1986turbulent,servidio2009magnetic,servidio2011magnetic,franci2017magnetic}.

On the other hand, magnetic reconnection can act
as a trigger of plasma turbulence in the reconnection outflows 
\citep{matthaeus1986turbulent,malara1991magnetic,malara1992competition,
lapenta2008self,huang2010scaling,beresnyak2016three}.
Numerical simulations have shown that reconnection outflows
can comprise different types of plasma instabilities at
both fluid \citep{guo2014rayleigh} and kinetic scales \citep{vapirev2013formation,huang2015magnetic}. 
These instabilities, along with the presence of  
shears, make reconnection outflows turbulent rather than laminar. 
The latter is not just a numerical evidence, but has been 
recently confirmed by in situ spacecraft observations \citep{eastwood2009observations,osman2015multi}.
Contrary to the ``classical'' view of reconnection energetics,
where the conversion of magnetic into kinetic energy happens 
only in the main reconnection sites \citep{shay2007two}, reconnection outflows are now also 
believed to be regions where the plasma is heated 
and particles are accelerated \citep{daughton2011role,lapenta2014electromagnetic,lapenta2015secondary}. 
It has been shown, both in numerical simulations \citep{leonardis2013identification,pucci2017properties}
and from most recent in-situ observation \citep{fu2017intermittent}, that, 
due to turbulence, energy exchange between fields and particles 
in the outflows is intermittent. This means that the greatest
part of energy transfer happens in low volume-filling,
very intense current sheets.
Recently, numerical simulations \citep{olshevsky2016magnetic} and observations \citep{fu2017intermittent} have shown that
energy exchange is more efficient in the presence of 
magnetic null points of spiral topological type (O-point configurations) rather than the radial nulls (X-points). 
This suggests than energy dissipation is more efficient in 
a region surrounded by two, or multiple, X-lines. 
In the classical two-dimensional MHD picture of magnetic reconnection, 
an O-point is always present between the two X-points (for topological reasons), 
i.e., in the middle of two counter-propagating reconnection jets, if the X-lines are active. 

The physics of reconnection jets collision has been
studied numerically with both fluid and kinetic models
\citep{karimabadi1999magnetic,nakamura2010interaction,oka2008magnetic,markidis2012collisionless},
sometimes in the frame of the tearing or plasmoid instability \citep{bhattacharjee2009fast,landi2015resistive}.
Recently, the first observations of colliding reconnection jets
have been carried out in the presence of both weak
\citep{alexandrova2016two} and strong \citep{oieroset2016mms} guide fields. 
Observational results show that jets collisions
cause wave activity and are responsible for secondary reconnection
events in the outflows, but a clear theoretical picture of reconnection jets collision
is still lacking. 
In this work we present a three-dimensional (3D) numerical simulation to study the
physics of reconnection jets collision, in the presence
of a strong guide field, at kinetic scales.
We show how jets collision gives rise to plasma turbulence
in a confined region of the outflows. 
We provide a numerical method to detect the turbulent 
region and perform local statistical analysis to characterize
the turbulence that is produced in such region. 
We believe that these results can be used for the interpretation
of present and future observations of jets collisions at ion 
and sub-ion scales. Moreover, we provide a theoretical 
description of turbulence generated by magnetic reconnection
in a three-dimensional kinetic model.

\section{Numerical set up} \label{sec:sec1}
We consider a plasma made of ions (protons) and electrons.
At the initial time the plasma is at the Harris equilibrium
$${\bf B} = B_{0x} \tanh(y/\delta) {\bf e_x} + B_{0z}{\bf e_z}, \;\;
  n=n_b+\frac{n_0}{\cosh^2(y/\delta)},$$
where $B_{0z}$ is the guide field, $\delta$ is the thickness of the Harris sheet,
and $n_{0b}$ the background density.
The coordinates are chosen such that the equilibrium sheared magnetic field 
is directed along the $x$ axis, and varies along $y$, causing the initial
current sheets to be directed parallel to the $z$ axis, which is also 
the guide field direction. 
For both species the initial distribution function is a Maxwellian 
with homogeneous temperature. 
The system obeys to the Vlasov-Maxwell equations, which are numerically 
solved using the semi-implicit particle-in-cell (PIC) code iPIC3D 
\citep{brackbill1982implicit,markidis2010multi,lapenta2012particle}.
The numerical domain is a Cartesian box with size $[40,\;30,\;10]\;d_i$,
where $d_i$ is the ion inertial length. We use $[512,\; 384,\; 128]$ cells,
each one initially populated with $125$ particles of each specie. 
A reduced ion/electron mass ratio $m_i/m_e=256$ is used, which fixes the spatial resolution to 
$\Delta x = 1.25 d_e$, where $d_e$ is the electron inertial length. 
The initial ion/electron temperature ratio is $T_i/T_e = 5$, 
and thermal velocities $u_{th,i}=0.0063c$ and $u_{th,e}=0.045c$, where $c$ is the speed of light. 
The thickness of the initial current sheet is 
$\delta = 0.5 d_i$ and the background density is $n_{b0}/n_0=1/10$. 
The value of the guide field is set to $B_{0z}/B_{0x}=2$.
We use open boundary condition in the $y$ direction and we impose
periodicity along $x$ and $z$. The equilibrium is perturbed by 
a Gaussian fluctuation of the $z$ component of the 
vector potential, located at ($x=0,\; y=7.5$), which initializes
the magnetic reconnection process \citep{lapenta2010scales}. The system evolution 
is followed in time up to $29 \tau_i$, where $\tau_i$ is the ion 
gyro-period, using a time step of $dt \approx \tau_e/10$, 
where $\tau_e$ is the electron gyro-period.  

\section{Dynamical evolution} 
\label{sec:sec2}
\begin{figure}
\includegraphics[width=\textwidth]{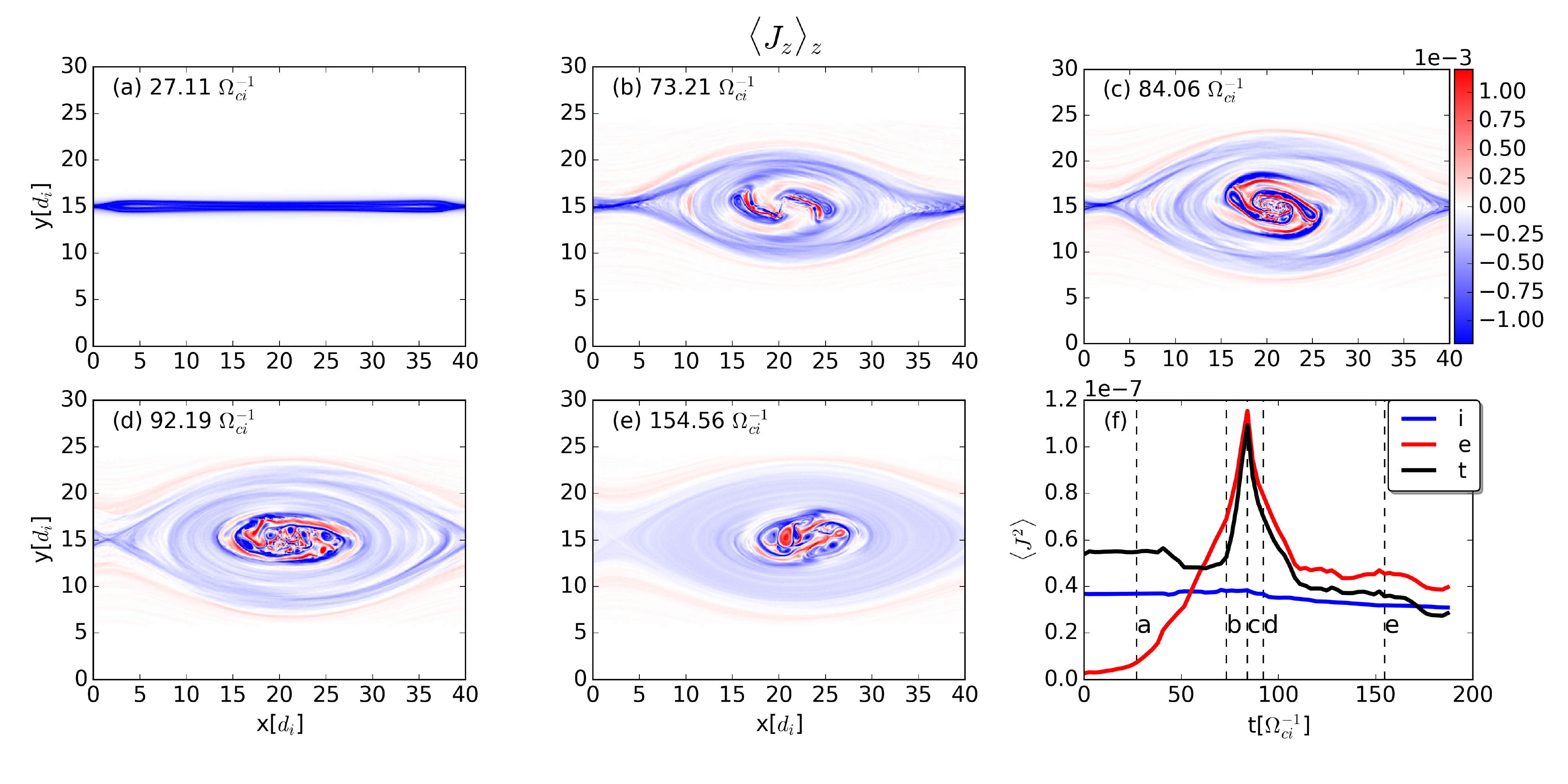}
\caption{(a-e) $J_z$ averaged along z at different times. (f) Mean square ion 
(blue), electron (red), and total (black) current as a function of time. Dashed
lines in panel (f) indicate plotting times of panels (a)-(e).}
\label{fig1}
\end{figure}

Because of the strong guide field, we expect that current sheets will be
preferentially aligned along the $z$ direction. In Figure \ref{fig1}, we represent a time 
evolution of the reconnecting current layer. The quantity plotted in panels (a-e) is $\langle J_z \rangle_z$, 
the out of plane total current density as a function
of $x$ and $y$ and averaged along $z$. The initial Harris
sheet is a narrow layer, directed along $z$, as reported in panel (a), and it
has been perturbed at the boundary $x=0$. Because of periodicity, two reconnection
sites form at opposite sides of the domain in the $x$ direction.
Panel (f) shows the evolution of the mean square ion, 
electron and total current $\langle J^2 \rangle$.
At the beginning, the current is mainly carried by ions, due to the Harris
equilibrium, but as the evolution
proceeds electron current becomes dominant. The total current reaches a peak 
value at $84.06 \Omega_{ci}^{-1}$ and rapidly decreases to a smaller 
value. Then, the value of $\langle J^2 \rangle$ remains almost constant,
slightly decreasing up to the end of the simulation.

The X-lines produce two reconnection jets which approach one against the 
other in the middle of the simulation domain (panel (b)). Notice that the 
two outflows carry oppositely directed current sheets even before the 
actual collision. When the two outflows collide, the maximum of $\langle J^2 \rangle$
is reached (panel (c)). At that instant the two outflows can still be 
distinguished even though
they look caught into each other. The two counter-propagating
outflows carry with them also oppositely directed magnetic fields, which makes
the system unstable to secondary reconnection. The large scale current sheets
present in panel (c), whose dimension are of the order of $\sim 5-10 d_i$, are
fragmented in smaller ones by multiple reconnection events. The effect is visible in
panel (d), where a multitude of current sheets at different scale
appears in the collision region. This suggests the beginning of a turbulent
cascade, that through nonlinear interactions forms smaller
current sheets. At later times, the turbulence which develops
in the center of the domain continuously generates 
and disrupts current sheets and the associated magnetic islands. 
It is worth noting that the area where the current sheets are 
located decreases in time (panel (e)). This is reasonable considering that the 
system is not driven externally, but is just relaxing after the 
jets collision. 

\begin{figure}
\includegraphics[width=\textwidth]{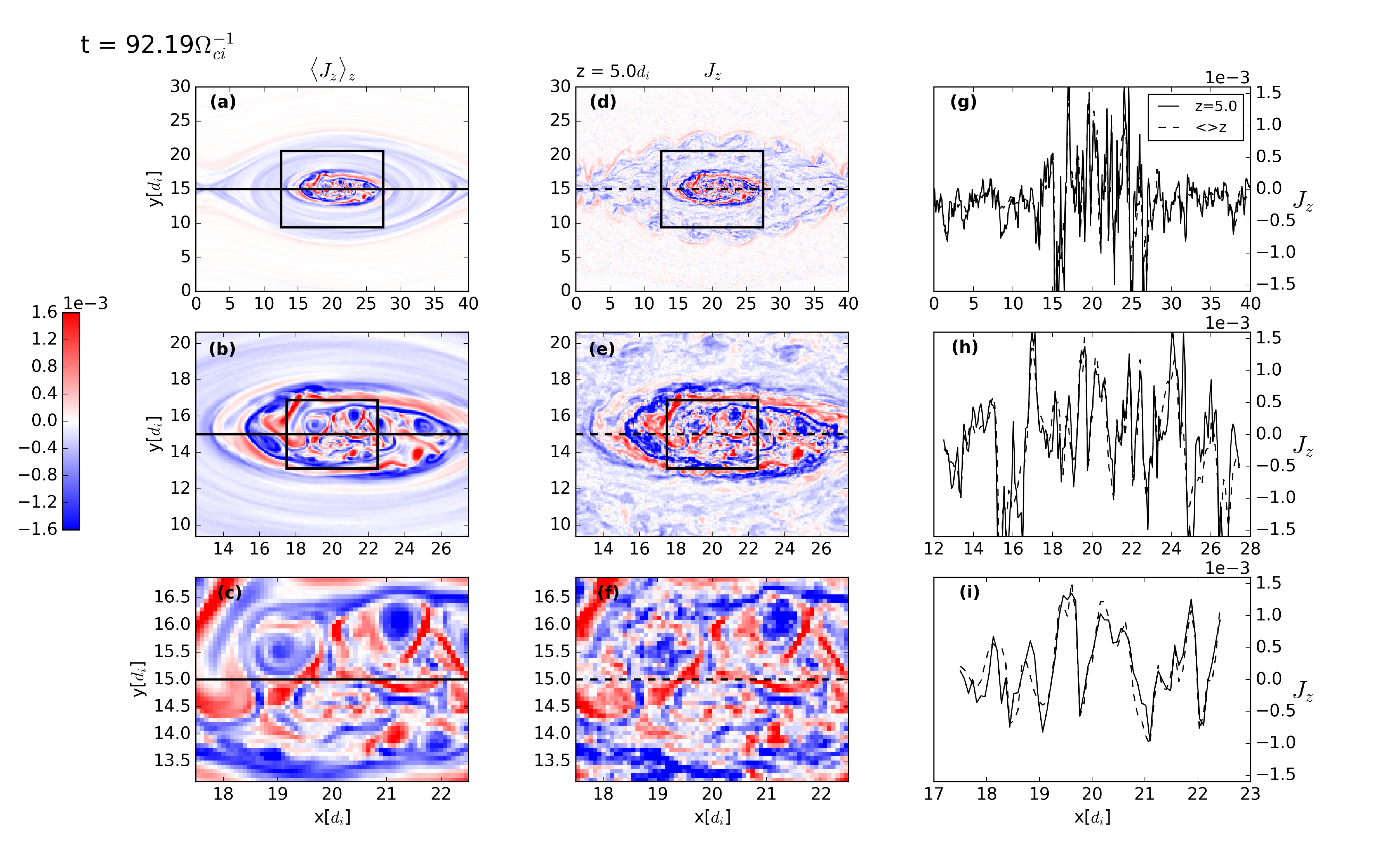}
\caption{Telescopic view of $J_z$ at $t=92.19 \Omega_{ci}^{-1}$. (a) Average out
of plane current density $\langle J_z \rangle_z$ and two zooms (b)-(c). (d) Out 
of plane current density $J_z$ at $z=5.0d_i$ and two zooms (e)-(f). (g) Current 
density profile as a function of $x$, at $y=15.0d_i$, for 
$\langle J_z \rangle_z$ (solid line) and for $J_z(z=5.0)$ (dashed line) and two 
zooms (h)-(i). The black frames in panels (a)-(b) and (d)-(e) bound the zoomed 
regions showed in panels (b)-(c) and (e)-(f), respectively. The solid or dashed 
black lines in panels from (a) to (f) identify the locations along which the 
profiles in panels (g)-(h)-(i) are plotted.}
\label{fig2}
\end{figure}

We have just presented a 2D evolution of the system, relying on the 
fact that the presence of a strong magnetic field favors the formation
of perpendicular gradients. To verify this assumption, in Figure \ref{fig2} we present a comparison 
of $\langle J_z \rangle_z$ with $J_z$ at a fixed $z$, namely $z=5.0 d_i$.
The plot is taken in the turbulent phase at $t=92.19 \Omega_{ci}^{-1}$ 
(panel (d) in Figure \ref{fig1}), right after the jets collision.
$\langle J_z \rangle_z$ and $\langle J_z \rangle(z=5.0)$ in the $xy$ plane
are plotted in the first and second column, respectively. The latter
looks like the blurry version of the former. This can be 
interpreted as follows. The current sheets are elongated in the $z$ 
direction and, for this reason, they are not ruled out by the average. 
At the same time, a small level of fluctuation with wavevector in the
$z$ direction are present, which make the picture of the 2D cut blurry.
Those fluctuations are ruled out when averaging along $z$. 
In the third column a 1D cut of the two quantities is plotted as 
a function of $x$. We can see that several current sheets are encountered
moving in the $x$ direction. The strongest ones are found in the central region. 
There is no big difference between the two variables, confirming that
the fluctuations in the $z$ direction are small compared to the average
current intensity. 
In the second and third row we show two zooms of the two quantities taken
in the central part of the simulation. We observe that the two quantities
look similar also at smaller scales. Moreover, we can notice that
the size of the smallest current sheets are of a fraction ($1/5$ or less) 
of the ion inertial length $d_i$, corresponding to $\sim 3 d_e$,
where $d_e$ is the electron inertial length.  

\begin{figure}
\begin{center}
\includegraphics[width=0.7\textwidth]{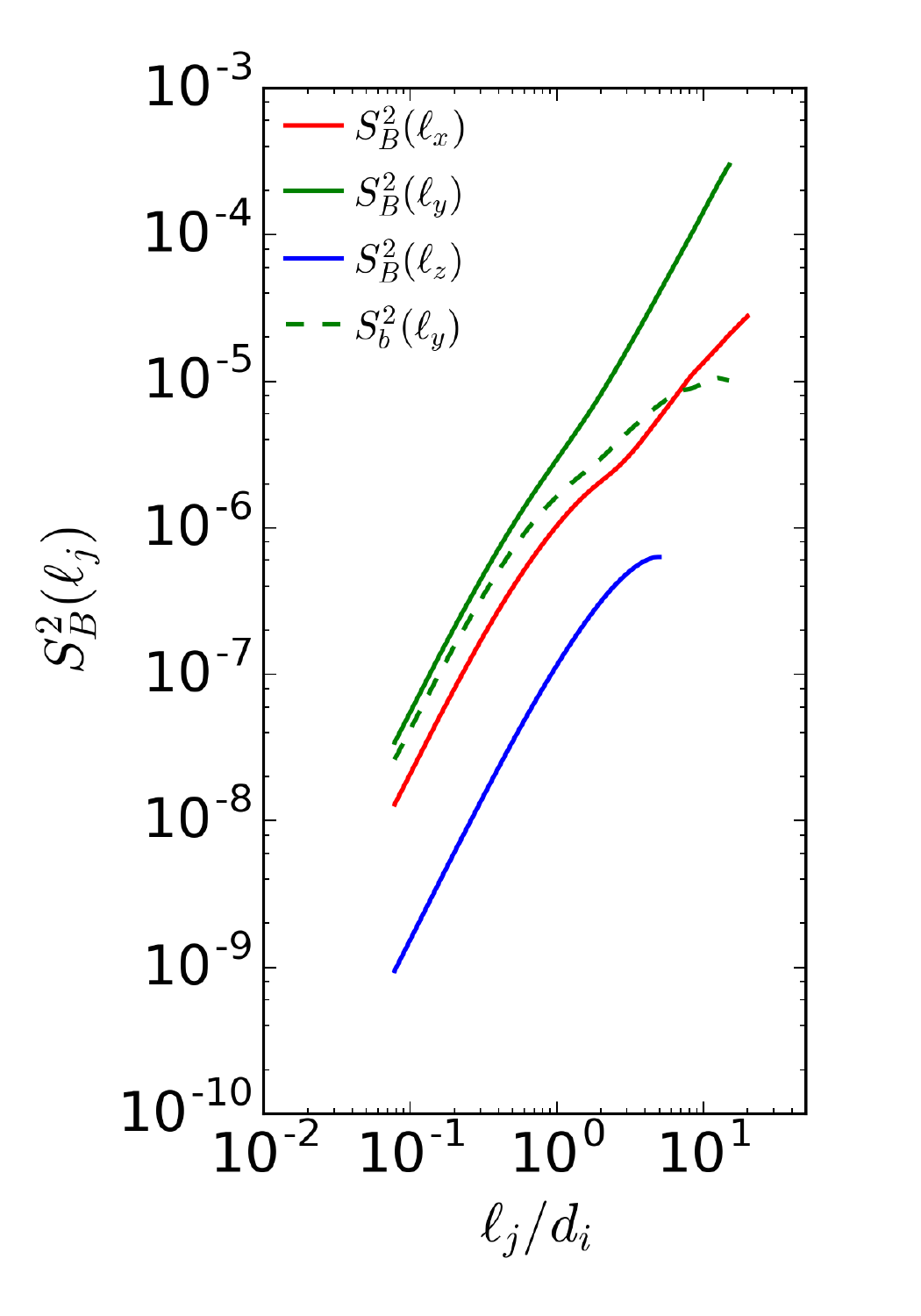}
\end{center}
\caption{Second order structure function of the magnetic field $S^2_B$ at 
$t=92.19\Omega_{ci}^{-1}$ for lags in the $x$ (red), $y$ (green solid) 
and $z$ (blue) direction. The green dashed line represents the second order 
structure function of the magnetic field fluctuations $S^2_b$ along $y$ for 
which the background Harris field has been subtracted 
from the actual field.}
\label{fig3}
\end{figure}

This preliminary analysis has shown that a turbulent behavior is 
observed after the collision of the two jets. The comparison
between $\langle J_z \rangle_z$ and $J_z (z=5.0)$ suggests 
that strong fluctuations are produced
perpendicularly to the guide field, and small amplitude fluctuations are 
present along the guide field. In order to make the last statement more 
quantitative, we have analyzed the second order structure function of the 
magnetic field. This is defined as 
$S_B^2({\bf r})= \frac{1}{V}\int |{\bf B}({\bf x}+{\bf r})-{\bf B}({\bf x})|^2 d^3x$,
where ${\bf B}$ is the total magnetic field, ${\bf r}$ is a displacement in the 
physical space, and $V$ is the total volume of the simulation box. 
Similar to the power spectral density, this quantity represents the average of 
the magnetic fluctuations energy between two points of the domain separated by a
lag ${\bf r}$. Since we are interested in evaluating the energy of the 
fluctuations varying in the direction parallel and perpendicular to $z$, i.e. 
the guide field direction, we have computed the so called reduced structure 
function defined as $S_B^2(\ell_i)= \frac{1}{V}\int |{\bf B}({\bf x}+\ell_i{\bf e}_i)-{\bf B}({\bf x})|^2 d^3x$,
with $i = x,y,z$. These quantities, plotted in Figure \ref{fig3}, represent 
the domain average of the magnetic energy fluctuations with wavevector in the 
$i$ direction at a scale $\ell$. For statistical reasons, we consider maximum 
lags $\ell_i^{max}=L_j/2$, where $L_j$ is the box size in the $j$ direction. 
The minimum lag is set to $\ell_j^{min} = \Delta x_j = 1.25 d_e$. 
The results confirm what suggested by the visual analysis of the current. 
The blue curve, representing $S_B^2(\ell_z)$, is always below the red and green 
curves, representing $S_B^2(\ell_x)$ and $S_B^2(\ell_y)$, respectively.
The magnetic fluctuation energy is larger for wavevector perpendicular to the 
guide field and smaller for parallel one at all scales. 
Moreover, the fluctuations in the direction of the initial shear, 
i.e. along $y$, are slightly stronger than the fluctuations in the $x$ 
direction. In order to check if this difference is due to the presence of the 
background Harris field we subtract its contribution 
to the total magnetic field, defining the new following variable:
${\bf b}={\bf B}-\langle {\bf B}_x(y)\rangle_{x,z}{\bf e}_x - B_{0z}{\bf e}_z$.
The variable ${\bf b}$ can be seen as the magnetic fluctuation field. 
$S_b^2(\ell_y)$ is plotted in Figure \ref{fig3} as a green dashed line. The 
fluctuation energy in the $y$ direction are stronger than in the $x$ 
direction even when the contribution of the Harris field is not 
taken into account. Both are at least one order of magnitude stronger 
than the fluctuations in the $z$ direction. We can also notice 
that at large scales the structure function tends 
to saturate only in the $z$ direction. The saturation is reached 
at the scale of the correlation length. Since large scale shears are present
in both the $x$ and $y$ direction, the corresponding structure function 
does not saturate for lags as large as half the box size in each direction. 
Consistently, when the contribution of the Harris field is subtracted
from the total field, the structure function in the $y$ direction 
saturates at around $5 d_i$. This size is comparable to the size 
of the turbulent region depicted in Figure \ref{fig1} (panel (d)) and Figure 
\ref{fig2}.

The analysis of $S_B^2$ conducts to the result 
that turbulence develops in the box after jets collision. This turbulence
is mainly two dimensional, with fluctuation energy concentrated 
in wavevectors perpendicular to the reconnection guide field. A second smaller 
anisotropy is found for the in plane fluctuations, where more energy is present 
in the direction of the initial magnetic shear.
The analyses performed so far are global, in the sense that the full 
computational domain was used as a single sample. However, the turbulent region 
seems to be confined in a smaller central region surrounding the location where 
the outflows have collided. In the next section we provide a method to isolate 
that turbulent region and study its properties separately from the rest of the 
domain.

\section{Isolating the turbulent region: method and local analysis}

In this section we provide a method for isolating the 
turbulent region and we describe the turbulence analysis 
performed on the region itself comparing it with the rest 
of the simulation box. 

\subsection{The Method} 
\label{sec:method}
In the previous section we have shown that turbulence develops
in the simulated reconnection event, due to jets collision, and seems 
to be localized in the surrounding of the collision site.
We have also shown that such turbulence is quasi two dimensional 
and develops in the plane perpendicular to the guide field. 
We build our method bearing on this last feature and considering
the system as being two dimensional, thus neglecting the dependence
on the $z$ component. 
In 2D MHD, the in-plane components of the magnetic field is 
a function only of the out-of-plane component of the 
vector potential ${\bf A}$. From ${\bf B} = \nabla \times {\bf A}$,
one finds that $B_x = \frac{\partial A_z}{\partial y}$, and 
$B_y = - \frac{\partial A_z}{\partial x}$, where $xy$ is the 
2D plane and $A_z$ is the $z$ component of ${\bf A}$. From the last
two relations it is easy to show that $({\bf B} \cdot \nabla)A_z=0$.
The last expression means that $A_z$ is constant moving along an 
in-plane field line or, equivalently, that in-plane magnetic field lines are 
isocontours of $A_z$. It is also easy to show that 
\begin{equation}\label{eq:laplA}
- (\nabla \times {\bf B})_z = \nabla^2 A_z.
\end{equation}
This equation will be used later.

\begin{figure}
\begin{center}
\includegraphics[width=\textwidth]{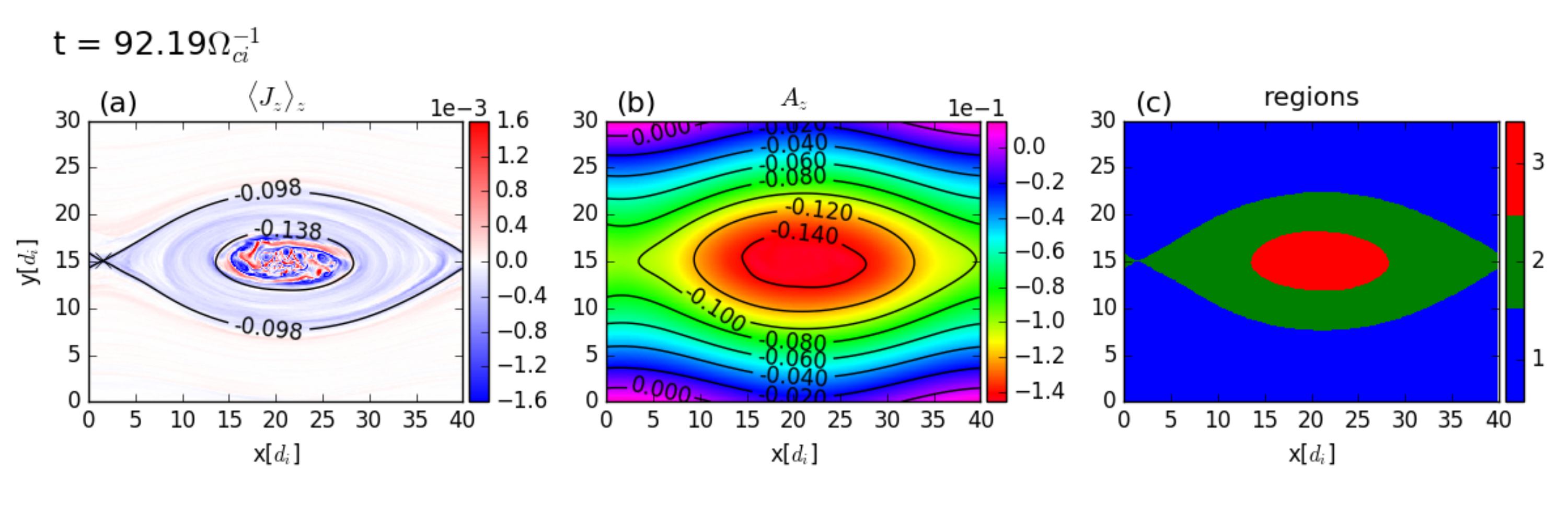}
\end{center}
\caption{(a) Out of plane current averaged along z $\langle J_z \rangle_z$ and 
isocontours of the out of plane component of the vector potential $A_z$, 
(b) $A_z$ and its isocontours, 
(c) three regions: R1 (blue), R2 (green), R3 (red), as described in 
section~\ref{sec:method}.}
\label{fig4}
\end{figure}

Operationally, the two dimensional ingredient 
is obtained by averaging the magnetic field $\bf B$ in space,
over the z direction. 
The z-average of the 2D magnetic field is similarly
related to the z-average of the full 3D vector potential, that is the 
2D potential $A_z = \hat z \cdot \langle {\bf A} \rangle_z$.
This averaging 
operation has the effect of eliminating the fluctuations in the $z$
direction, which are however much smaller than those in $x$ and $y$,
as shown previously. 
The map of the 2D potential 
$A_z$ is plotted
in panel (b) of Figure \ref{fig4}, along 
with its isocontours, which are the in-plane magnetic field lines. 
In order to compute $A_z$, we solved Equation (\ref{eq:laplA}) 
using a Fourier method in an expanded (periodic) domain, 
then plotting the result in the original domain. 
The magnetic island produced by reconnection in the 
center of the box is clearly visible. Moreover, we notice that $A_z$ has a 
minimum in the middle of the island and a saddle point at the location of the 
main X-line. This is consistent with what found in incompressible MHD 
simulations of 2D magnetic reconnection \citep{matthaeus1982reconnection}.
It is also consistent with Equation (\ref{eq:laplA}), as we explain in the 
following. The presence of a minimum implies that $\nabla^2 A_z > 0$, and 
consequently $(\nabla \times {\bf B})_z < 0 $. From Maxwell equations, 
we know that $\nabla \times {\bf B} = \frac{1}{c}\frac{\partial {\bf E}}{\partial t} + \frac{4\pi}{c}{\bf J}$.
At large scales and slow frequencies, the displacement current term can 
be neglected and the curl of ${\bf B}$ becomes proportional to ${\bf J}$.
Considering that current in the initial current sheet is negative for 
construction, the minimum of $A_z$ in the middle of the island is justified.
In panel (a) of Figure \ref{fig4} we plot a 2D map of $\langle J_z \rangle_z$ 
at time $t=92.19 \Omega_{ci}^{-1}$ and two isocontours of $A_z$, i.e. 
$A_z = -0.098$ and $A_z=-0.138$. The former represents the value of $A_z$ at the 
X-point and at the separatrices, the latter a manually selected value of $A_z$ 
associated with a magnetic field line enclosing most of the 
more turbulent central region and very little of the more 
quiescent (low current density) region surrounding it.   
These two isocontours identify three different regions, plotted in panel (c): 
Region 1 (R1), out the separatrices; Region 2 (R2), between the separatrices
and the turbulent region; Region 3 (R3), the central turbulent region. 

Since we want to reproduce such a distinction for each instant of time 
we have implemented the following numerical procedure. First, at a given time 
$t$ we compute the in-plane magnetic field average along $z$ and the 
corresponding 2D vector potential. Second, we find $A^X_z$, namely the value of 
$A_z$ at the X-point, and $A^{min}_z$, namely the minimum of $A_z$ located in 
the center of the domain. We call $A^*_z$ the suitable value of $A_z$ associated
to the magnetic field line that encases the turbulent region. Since this value
has to lay between $A^{min}_z$ and $A^X_z$, we can write it in this form:
\begin{equation}
 A^*_z = A^{min}_z + a(A^X_z-A^{min}_z), {\mbox {with }} 0 < a < 1.
\end{equation}
Third, we find the suitable value of $a$ as that associated to a contour
that encases all the strongest current sheets in the turbulent region. We found
$a=1/6$ to be suitable for time $t=92.19\Omega_{ci}^{-1}$. Then, we repeat the 
procedure for each time step, keeping the selected value of $a$ unchanged. The 
procedure is applied from the time when the mean square current 
is maximum, i.e. $t=84.06\Omega_{ci}^{-1}$, up to the end of the simulation. 
Before that time it has no meaning separating the domain in regions, 
since turbulence has not developed yet.
Choosing a fixed value for $a$ does not mean in principle that $A^*_z$ does not 
change in time, since both $A^{min}_z$ and $A^X_z$ can change. It rather means 
that we are choosing a value of the vector potential whose relative distance
from the minimum and the value at the separatrices is constant. 
Interestingly, this choice gives a correct result for all time steps 
(see attached video). This can be explained considering that, from Maxwell's 
equations, $\frac{\partial A_z}{\partial t}=-cE_z$, meaning
that the value of $A_z$ changes due to the reconnection electric field.
After the initial phase when jets are ejected, the value of this field 
become smaller causing the vector potential to be nearly constant in time on the 
separatrices. 
This method allows to identify different regions in the simulation domain in 2D. 
Since small fluctuations are present in the $z$ direction, we can extend the 
2D region along $z$ and perform the statistical analysis on the 3D 
extension of each region.

\subsection{The Analysis}

\begin{figure}
\begin{center}
\includegraphics[width=\textwidth]{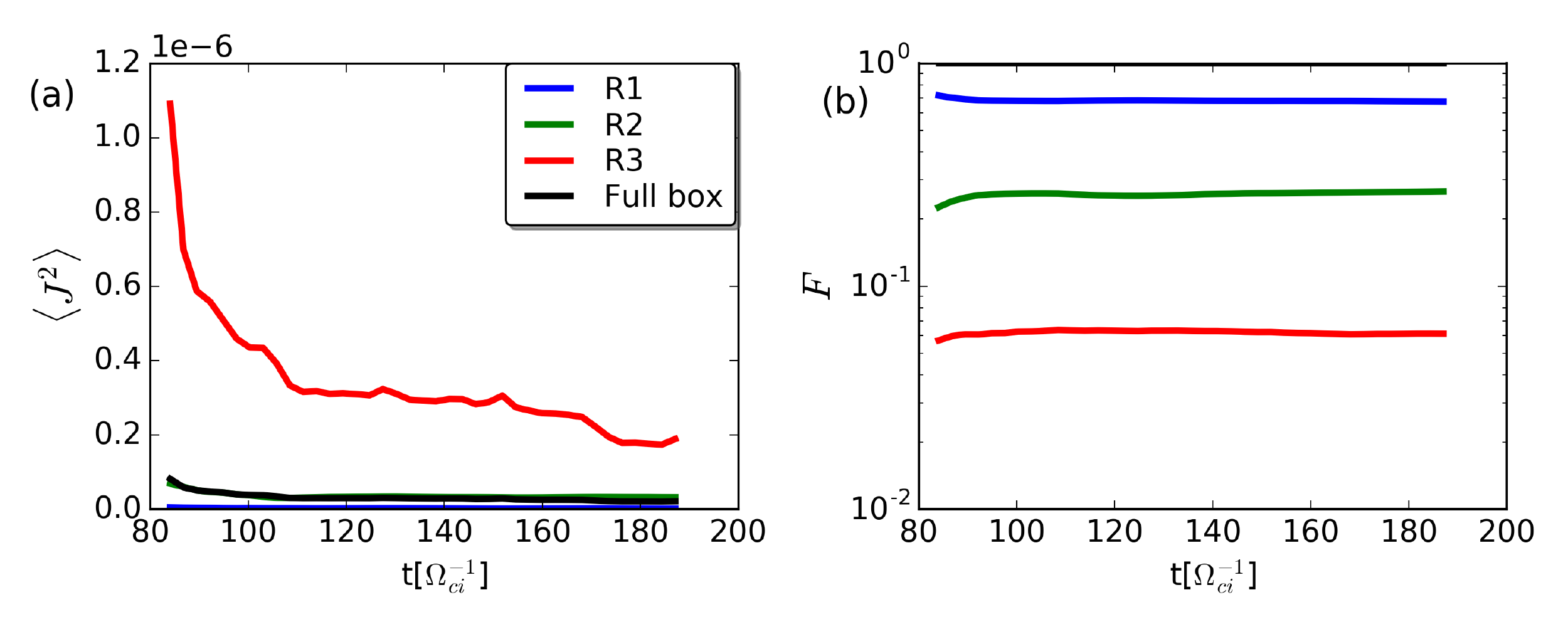}
\end{center} 
\caption{(a) Mean square current $\langle J^2 \rangle$ as a function of 
time and (b) filling factor $F$ computed as the ratio between the volume of a 
region and the volume of the full box, for R1 (red), R2 (green), 
R3 (blue), and full box (black).
}
\label{fig5}
\end{figure}

We have identified the turbulent region looking at the out of plane component
of the total current, thus associating turbulence with the current activity. 
In Figure \ref{fig5} we show the mean square current in the three regions R1, 
R2, and R3, as a function of time (panel (a)). The full box average is plotted 
as a reference. The mean square current in R3 is much higher than in R1 and R2. 
Its trend in time looks very similar to the trend of the full box average, 
shown in panel (f) of Figure \ref{fig1}.
Panel (b) shows the filling factor $F$ of each region as a function of time, 
computed as the ratio between the volume of each region and the volume of the 
whole box. R3 is the smallest region and occupies around 6\% of the total 
volume. The size of the region does not change significantly in time. This 
analysis shows that the most intense current sheets are within region R3, which 
is a small portion of the total domain. However, region R3 is large
enough to perform a statistical analysis of fluctuations contained therein.

\begin{figure}
\begin{center}
\includegraphics[width=\textwidth]{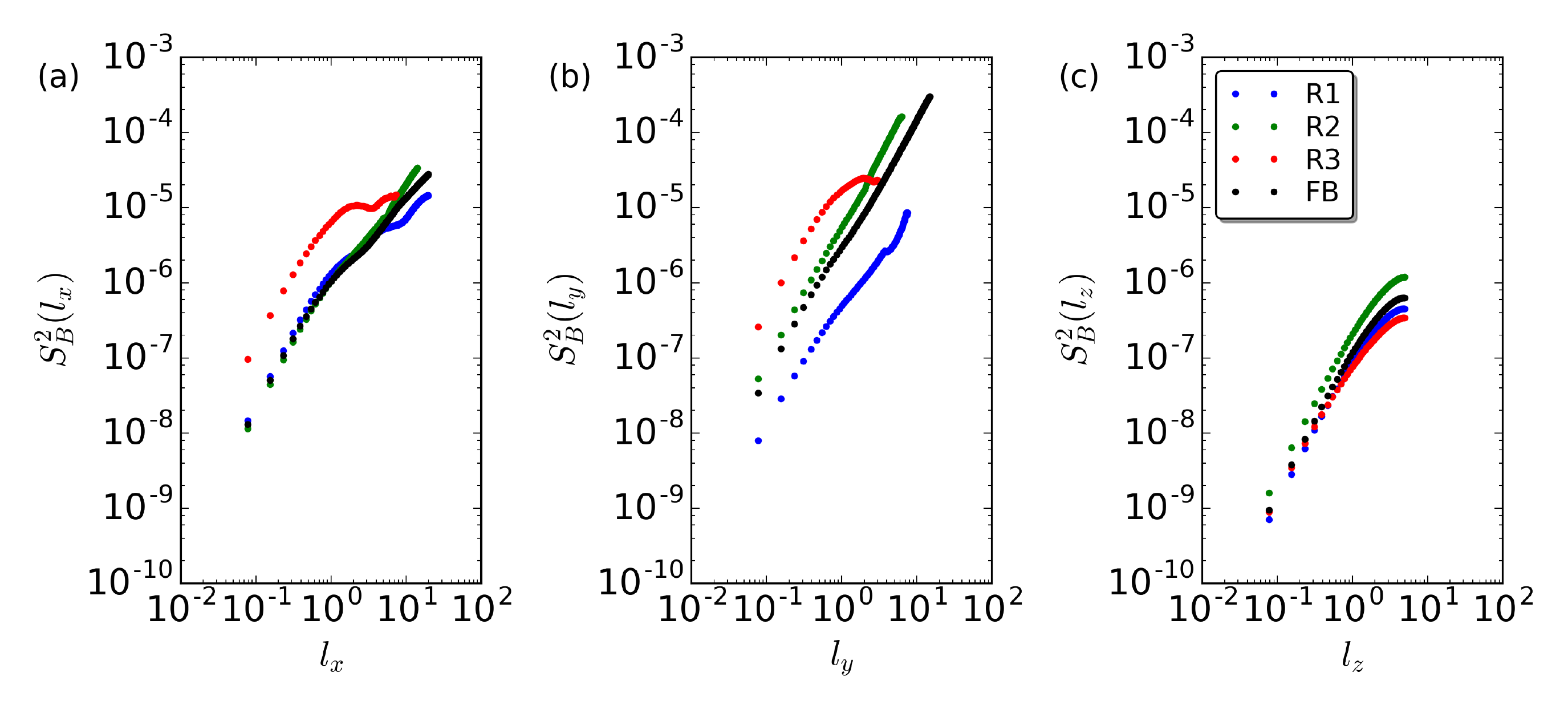}
\end{center} 
\caption{Second order structure functions of the magnetic field $S^2_B$ at 
$t=92.19 \Omega_{ci}^{-1}$ for lags $\ell$ along $x$ (a), $y$ (b) and $z$ (c). 
$S^2_B$ is computed in R1 (blue), R2 (green) and 
R3 (red), and in the full box (black).
}
\label{fig6}
\end{figure}

Since the chosen regions are not rectangular, and their boundaries are 
not periodic, a Fourier-based spectral analysis is not a suitable 
approach. For this reason, as done for the full 
box analysis, we compute the second order structure functions of the 
magnetic field in the three regions for lags in the $x$, $y$ or $z$ 
directions. In Figure \ref{fig6} we report the result at time 
$t=92.19\Omega_{ci}^{-1}$. We notice first that the energy for parallel 
increment lags (panel (c)) is smaller than for perpendicular lags (panels (a) 
and (b)), in each region (panel (c)). Moreover, different regions have similar 
parallel fluctuations energy at all scales. Panels (a) and (b) show the second 
order structure function of the in-plane magnetic fluctuations. Due to the 
difference in shape and size of the regions, the curves do not span the same 
range of lags.
The maximum possible lag is in fact defined as half of the sample size. The red
curve, associated to region R3, terminates at smaller lags because it is 
obtained in the smallest region R3. Moreover, the maximum $\ell_x$ for R3 is 
bigger than the maximum $\ell_y$, since the shape of the region is elongated in 
the $x$ direction. At large scales, a saturation in the value of the 
second order structure function is observed just for R3. The lag at  
which the saturation is observed gives an estimate of the correlation 
length of around $5 d_i$ along $x$ and $3 d_i$ along $y$, which is comparable to 
the sizes of the region. Such saturation is not observed for R1 and R2,
due to the presence of the large scale magnetic shear whose dimension is 
comparable to the sizes of those regions. At scales smaller than the correlation 
length, the red curve (R3) is always higher than green (R2) and blue (R1),
$S^2_{B,R3}>S^2_{B,R2}>S^2_{B,R1}$.

The average energy of magnetic fluctuations varying in the plane 
perpendicular to the guide field is larger in R3, implying that 
the non-linear cascade has acted more efficiently in that region. 
Therefore, this proves that R3 is the actual region where turbulence 
develops.

\begin{figure}
\begin{center}
\includegraphics[width=\textwidth]{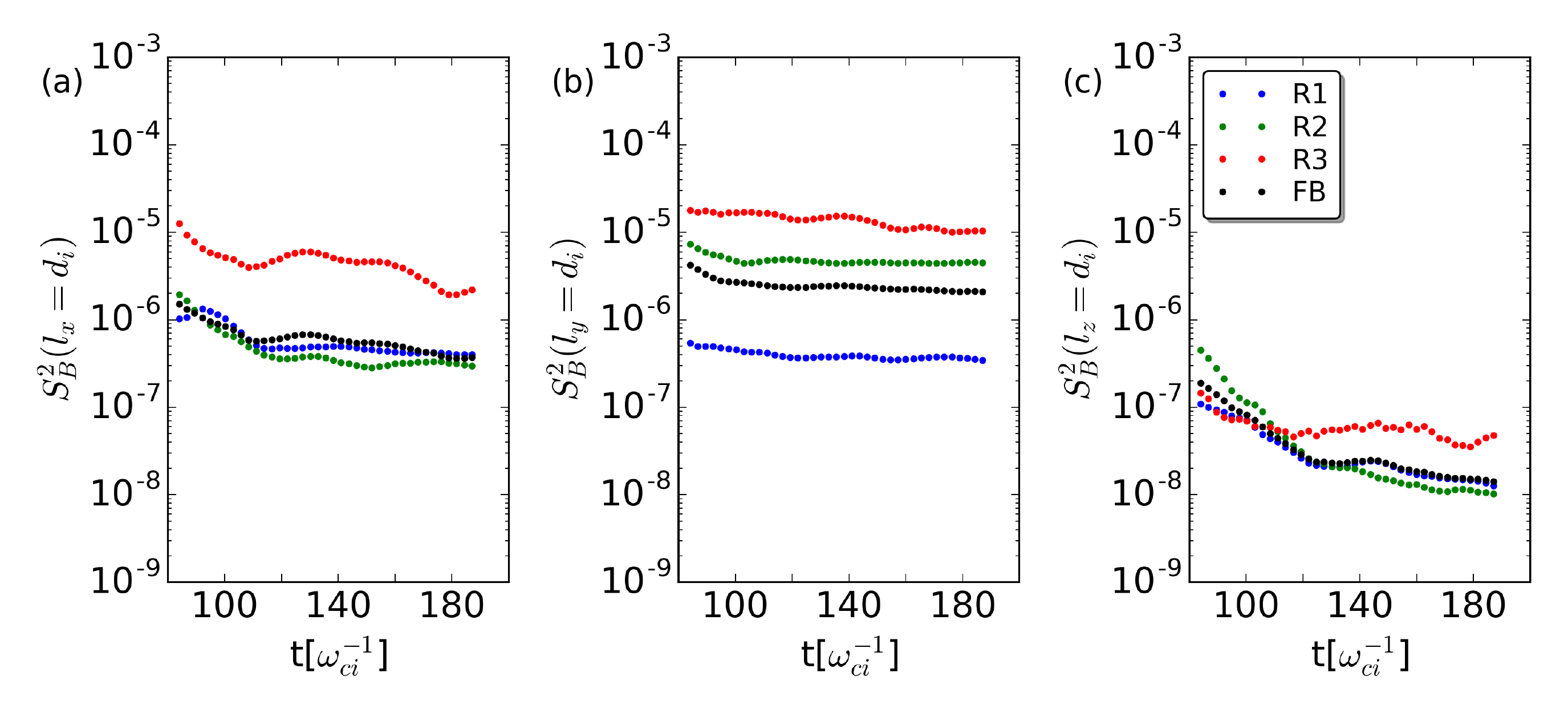}
\end{center} 
\caption{Second order structure functions of the magnetic field $S^2_B$ at 
$l_i= d_i$ as a function of time for $i$ equal to $x$ (a), $y$ (b) and $z$ (c). 
$S^2_B$ is computed in R1 (blue), R2 (green) and 
R3 (red), and in the full box (black). 
}
\label{fig7}
\end{figure}

In order to confirm that the turbulent cascade proceeds up to
the end of the simulation, we have computed $S^2_B$ in each 
region for different times. At each time we take the value
of $S^2_B$ for $\ell_j= d_i$ for $j=x,y,z$. 
The result is plotted in Figure \ref{fig7}. In the perpendicular
directions the average of the magnetic fluctuation energy at 
scale $d_i$ is larger in R3, meaning that more energy is 
cascading to sub-ion scales in R3 compared to the rest 
of the domain. The energy in the $y$ direction is larger 
on average due to the presence of the Harris field. 
In the $z$ direction the values in R3 become eventually 
larger as the simulation proceeds. At earlier times, 
the large energy is R2 is due to the formation of an electron 
instability that develops on the separatrices 
\citep{daughton2011role,fermo2012secondary}. 
This instability remains confined in R2 and 
saturates for times larger than $t \sim 100 \Omega_{ci}^{-1}$. 

To complete our analysis, we study the Eulerian frequency spectra of the 
electric and magnetic field in R3. The fields have been sampled during
the simulation at a cadence of $3\Delta t$ by means of probes 
(`virtual satellites') placed at predefined locations inside the simulation box. 
For the analysis we present here, we have used satellites from a region of 
dimensions $1.25 \times 1.25 \times 10.0 \; d^3_i$ located in the middle of R3. 
The selected region contains $3 \times 3 \times 24 = 216$ equally spaced 
satellites. We analyze the data in the time frame 
$92.18 \Omega_{ci}^{-1}<t<187.09\Omega_{ci}^{-1}$, extending from the initial 
development of the turbulence to the end of the simulation. Similar to what 
happens for in-situ spacecraft observations, magnetic field data are not 
periodic. Moreover, the situation we are considering here is not stationary, 
since turbulence in R3 is decaying in time. In order to avoid spurious effects 
coming from signal detrending or windowing, we use the technique proposed by 
\citet{bieber1993long} to compute the frequency power spectra. 
This method has been used, e.g., for the analysis of the recent in situ 
solar wind observations \citep{bale2005measurement}. 
The method follows these steps. First, the series of the increments
is computed from the actual field. If {\bf B}(t) is the magnetic field
as a function of time, its increments are defined as: 
$\Delta {\bf B}(t) = {\bf B}(t + \Delta t) - {\bf B}(t)$. 
Second, the autocorrelation function of the increments is computed as 
$R^{(\Delta)}_{ii}(\tau) = 
  \frac{1}{V}\int\Delta {\bf B}_i(t) \Delta{\bf B}_i(t+\tau)dt$.  
Third, from the autocorrelation function the power spectrum 
from the increments $P^{(\Delta)}_{ii}(f)$ is computed via Fourier transform.
Finally, the spectrum of the actual field $P_{ii}(f)$ is recovered by filtering 
the spectrum of the increments:
$$ P_{ii}(f) = \frac{P^{(\Delta)}_{ii}(f)}{4\sin^2(\pi/\Delta t)}.$$
This procedure suppresses the contribution 
of the low frequencies unresolved in the chosen sample.
Computing power spectral density from autocorrelation function can yield 
to non-physical results such as negative power spectrum values 
\citep{blackman1958measurement}, therefore we limit our analysis to lags smaller 
than $\tau_{max}=T/10$, where $T$ is the duration of the total sample.
With this upper limit we have enough statistics to estimate the spectrum 
for all frequencies selected.

\begin{figure}
\begin{center}
\includegraphics[width=\textwidth]{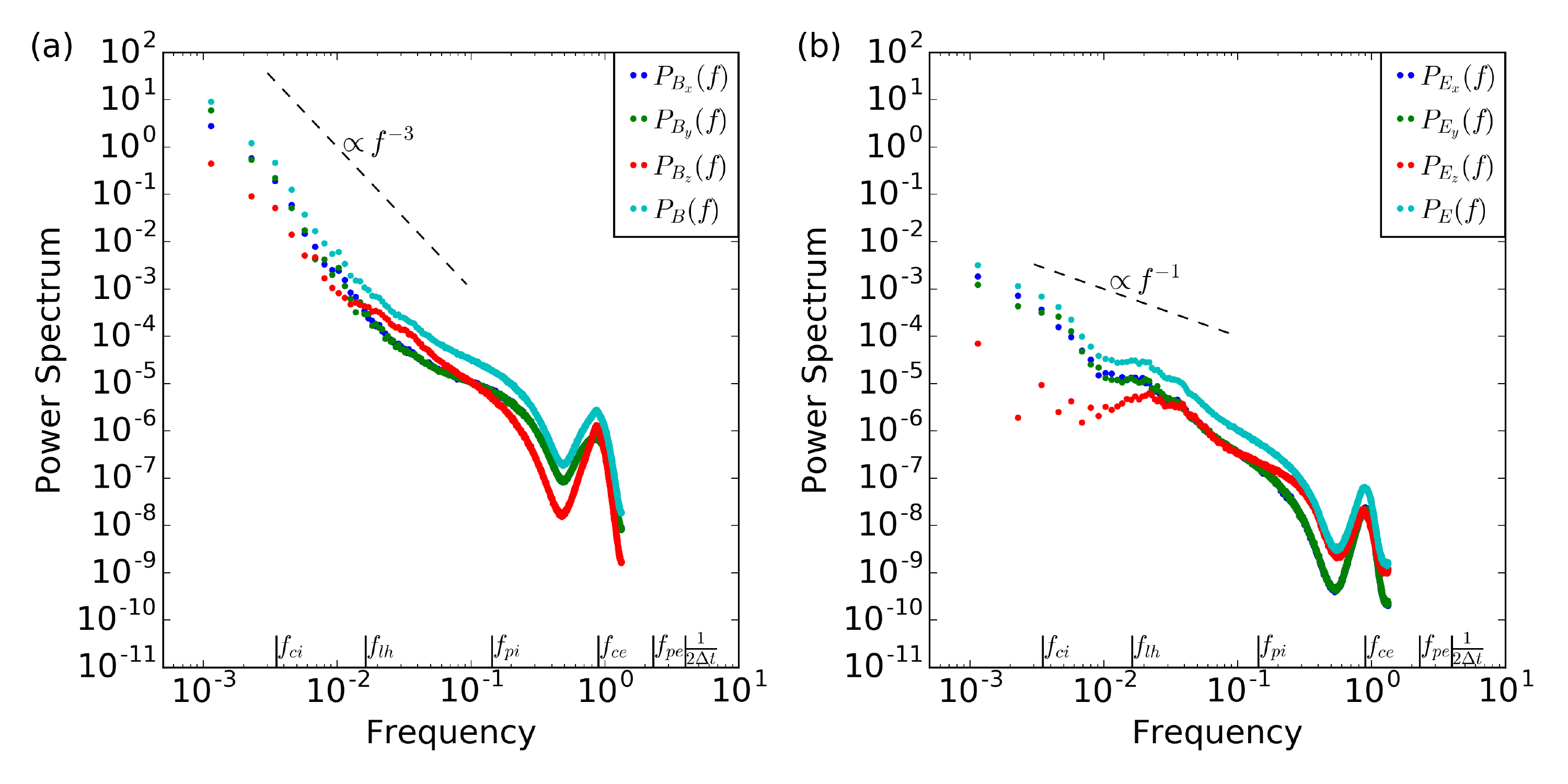}
\end{center} 
\caption{Power spectrum of magnetic (a) and electric (b) fluctuations for $x$ 
(blue), $y$ (green) and $z$ (red) component, and for the total field 
(light-blue). The black dashed lines represent a power law of the type 
$\propto f^{-3}$ in panel (a) and of the type $\propto f^{-1}$ in panel (b). 
In both panels vertical lines on the $x$ axis represent: ion cyclotron frequency 
$f_{ci}$, lower-hybrid frequency $f_{lh}$, ion plasma frequency $f_{pi}$, 
electron cyclotron frequency $f_{ce}$, electron plasma frequency $f_{pe}$, 
and the Nyquist frequency $1/ 2 \Delta t$, computed and averaged in the region 
where the probes are located.}
\label{fig8}
\end{figure}

Figure \ref{fig8}, reports the power spectra of magnetic (panel (a)) 
and electric (panel (b)) field as a function of frequency. The spectra are 
computed for each probe separately and then averaged. 
In Figure \ref{fig8} both 
spectra of the three components and total field are plotted. 
Characteristic frequencies are marked by vertical lines. 
Reference lines are shown 
correspnding to 
spectral indices of 
$-3$ and $-1$.
The computed spectra 
span frequency ranges 
that extend from a fraction of the 
ion cyclotron frequency to beyond the electron cyclotron frequency. The electron
plasma frequency is poorly resolved. Note that 
the range of plotted 
frequencies doesn't reach the Nyquist frequency because the fields are saved 
each $3 \Delta t$. Magnetic and electric frequency spectra from fluid to 
electron scales in plasma turbulence have been recently studied by means of 
in-situ spacecraft observations in the magnetosheath by 
\citet{matteini2016electric}. It has been shown that the magnetic and electric spectra
have approximately similar behavior in the fluid-MHD like regime, while they 
show different trends in the kinetic range. In particular, the magnetic field 
spectrum steepens at ion scales, passing from a spectral index between $-5/3$ 
and $-3/2$ in the MHD fluid range to a roughly $-3$ spectral index at ion and 
sub-ion scale. The 
electric field has an opposite trend, passing from a 
spectral index $-3/2$ in the fluid range to roughly $-1$ in the kinetic range. 
The familiar theoretical explanation for this phenomenon (e.g., 
\cite{MattEA08-comment}) depends on dominance in the kinetic range
of non-ideal terms in the generalized Ohm's law. 
When the turbulence $\bf k$-vectors are mainly 
perpendicular to the guide field, this heuristic explanation 
predicts $\delta E^2 \propto k^2\delta B^2$ for an electromagnetic fluctuation at a given wavevector $k$ in the sub-ion range.

The above reasoning 
explains the relationship of 
observed frequency and wavenumber spectra in cases in which the
two are proportional and 
related by the Taylor hypothesis, appropriate for high speed 
super-Alfv\'enic flows. 
This explains the agreement of frequency spectra coming from 
single spaceraft observations 
\citep{bale2005measurement,eastwood2009observations,matteini2016electric}, with 
wavenumber spectra obtained from numerical simulations 
\citep{pucci2017properties,franci2017magnetic}.

The results we report here seem to be, at a first sight, in contrast 
to the above scenario, in that 
the sub-ion scale 
magnetic and electric spectra in 
our analysis do not follow a specific power law 
and strongly differ from a $\propto f^{-3}$ and a $\propto f^{-1}$ trend, 
respectively. 
The main 
reason for this apparent discrepancy relies on the fact that Taylor hypothesis 
is not valid in the system considered here, as we presently demonstrate. 

The Taylor hypothesis consists in 
assuming that time variations at a single observational point
are only due to the rapid sweeping of spatial 
structure past that point. 
This requires that 
the large scale bulk flow of 
the plasma greatly exceeds 
all characteristic dynamical speeds that might distort
those structures. 
This includes both wave propagation 
and turbulent velocity \citep{perri2017numerical}. 
When Taylor hypothesis is not verified, there is 
not a simple connection between the frequency and wavenumber spectra, 
although there remains the possibility of 
a statistical similarity between the 
Eulerian frequency spectra and the wavenumber spectra 
\citep{chen1989sweeping,matthaeus1999dynamical} when the dominant effect
is random large 
scale sweeping.

\begin{figure}
\begin{center}
 \includegraphics[width=\textwidth]{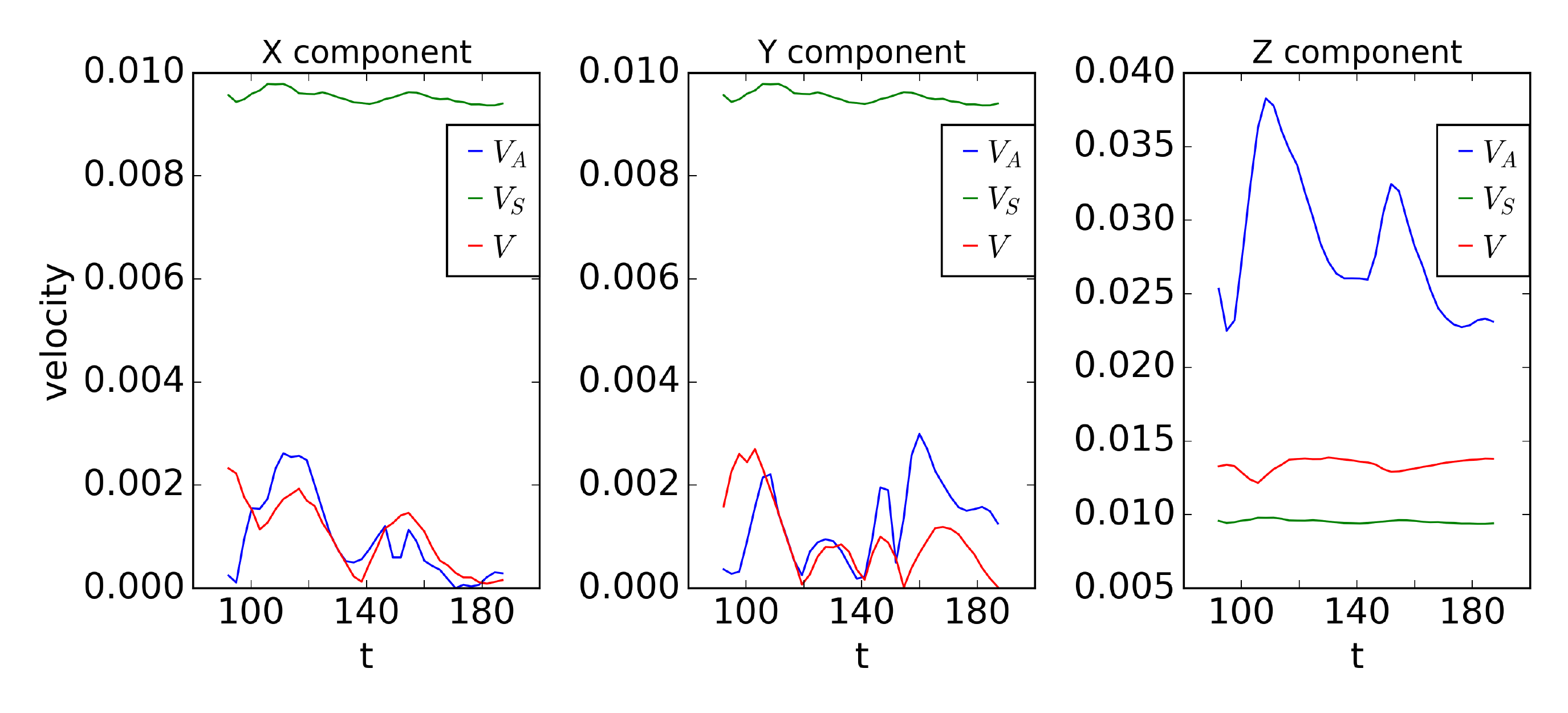}
\end{center} 
\caption{Characteristic speeds in the probes region. $V_s$, $\bf{V_a}$ and 
$\bf{V}$ are the average ion sound speed, the average Alfv\'{e}n velocity and 
the average bulk velocity, respectively. The three panels show the $x$ (left), 
$y$ (middle), and $z$ (right) components of the velocities and the ion 
sound speed.
}
\label{fig9}
\end{figure}

The Taylor hypothesis in not valid in the system
we are considering here. In order to assess this hypothesis, 
the bulk velocity of the 
plasma, computed in the region where the virtual satellites are located, must be
compared with characteristic wave speeds and with turbulent velocity 
fluctuations. In Figure \ref{fig9} we report the comparison between the bulk 
velocity ${\bf V}$, the Alfv\'en velocity ${\bf V}_A$, and the ion sound speed 
$V_s$. The bulk velocity ${\bf V}$ is defined as the plasma center of mass 
velocity, the Aflv\'en velocity as ${\bf V_A} = {\bf B}/\sqrt{4\pi\rho}$, where 
$\rho$ is the ion density, and the ion sound speed as 
$V_{S} = \sqrt{\frac{\gamma \sum_jP_{jj}}{3\rho}}$, where $j=(x,y,z)$, $P$
is the ion pressure tensor, and $\gamma$ is the adiabatic index, which we have 
chosen equal to $5/3$ for simplicity. 
These characteristic macroscopic speeds are compared in Figure \ref{fig9}, which 
shows the $x$, $y$ and $z$ components of ${\bf V}$ and ${\bf V}_A$, and the the ion sound speed
(considered isotropic). 
The bulk speed is in every case smaller or comparable with the other two characteristic speeds. In particular, the ion sound speed is the larger speed in
the directions perpendicular to the guide field, while the Alfv\'en speed is, as
expected, the largest speed in the parallel direction. 

\begin{figure}
\begin{center}
\includegraphics[width=\textwidth]{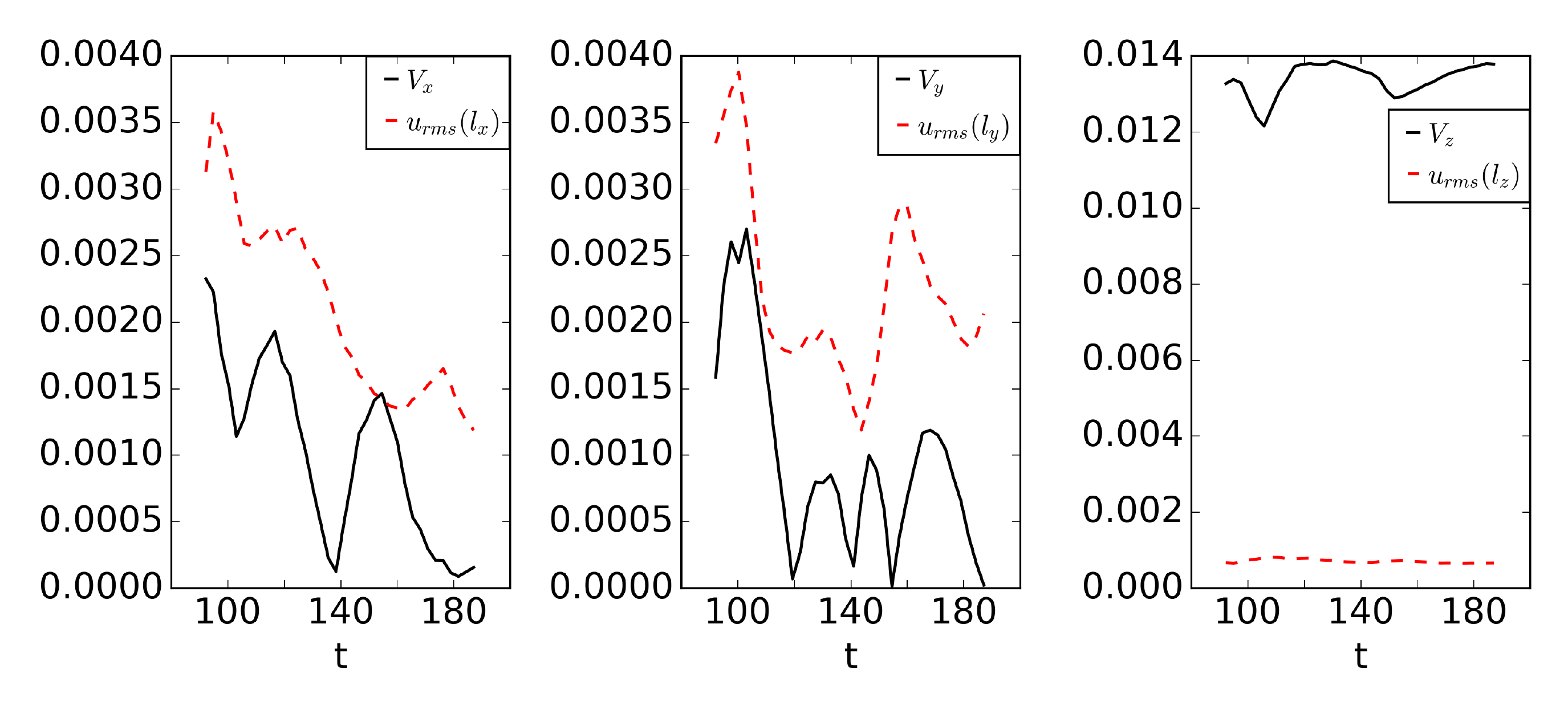}
\end{center} 
\caption{Comparison of bulk speed with turbulent velocity at scale $d_i$. The 
three panel show the $x$ (left), $y$ (middle) and $z$ (right) components of the 
velocities. The bulk speed components are plotted in black, while red dashed 
curves represent the turbulent velocity computed considering
lags directed along $x$, $y$, and $z$, respectively.}
\label{fig10}
\end{figure}

Next the bulk speed is compared with measures of 
the turbulent fluctuation velocities in Figure \ref{fig10}. 
In particular we compute the longitudinal increments 
${u}({\bf r},{\bf \ell}) = \hat{{\bf \ell}} \cdot [{\bf V}({\bf r} +{\bf \ell}) - {\bf V}({\bf r})]$, where ${\bf r}$ is a position in space, and ${\bf \ell}$ is the direction along which the velocity increments are computed. The root mean square velocity increments are computed 
in terms of components along the three Cartesian directions as $u_{rms}(\ell_i) = \sqrt{\langle {\bf u}({\bf r},{\bf \ell_i}) \cdot {\bf u}({\bf r},{\bf \ell_i}) \rangle}$. Here 
$\langle \rangle$ means average in the region where the probe are located, and $\ell_i$ with $i=x,y,z$ indicates the direction along which the velocity increments are computed. We considered a lag size equal to $d_i$ for each direction. In this way, we examine in particular whether the Taylor hypothesis is valid for the kinetic features seen in the frequency spectra in Figure \ref{fig8}. 
Figure \ref{fig10} shows a comparison of the bulk velocity with 
these measures of fluctuations. 
Each panel contains the i-th component of the bulk velocity and the corresponding $u_{rms}(\ell_i)$. First, we notice that $u_{rms}(\ell_z)$ is smaller than $u_{rms}(\ell_x)$, and $u_{rms}(\ell_y)$, in accordance with anisotropic, quasi 2D turbulence developed in the plane perpendicular to the guide field. 
Second, we notice that $V_z$ the bulk speed along the guide field direction is larger than all $u_{rms}$. This large speed 
could in principle be responsible for a similarity between the $k_z$ spectrum and the frequency spectrum. However, as shown in the previous section, turbulence 
is mainly 2D due to the presence of the strong guide field, and very little 
energy is contained in the $k_z$ modes. 
Moreover, the largest characteristic speed in the $z$ direction is the Alfv\'en speed (see panel (c) in Figure \ref{fig9}), which allows for the presence of fast parallel propagating waves of Alfv\`enic nature that can 
invalidate the Taylor hypothesis.

Based on this analysis of characteristic speeds, we can conclude that the Taylor hypothesis is not verified in the location where the Eulerian spectra are computed.
Therefore a similarity between frequency and wavenumber spectra is not expected.
The magnetic and electric frequency spectra presented 
in Figure \ref{fig8} are therefore to be considered as 
independent of the spatial (wavenumber) structure. 
Then the observed features are essentially temporal, 
and the observed features 
indicate the presence of wave activity at different temporal 
scales, which we now briefly describe.

First, we notice a flattening of the magnetic spectrum 
between the ion and electron cyclotron frequency, around $f=10^{-1}$, relative to 
the $\propto f^{-3}$ trend at lower frequencies. This feature is consistent with the presence of whistler waves and is also found in in-situ observations \citep{matteini2016electric}. To confirm this hypothesis, we applied a pass band filter at a frequency $f_{ce}/5$ to the magnetic field (not shown here) finding a phase shift between $B_x$ and $B_y$ consistent with parallel propagating whistler. 

Second, a significant peak at the electron cyclotron frequency is observed in 
both electric and magnetic field spectra. The peak is found for all the three 
components of the fields probably indicating the presence of both 
electromagnetic and electrostatic modes. 
To verify that this peak is not due to a numerical artifact, we ran 
a new simulation (not shown) with an halved time step and saving the fields data at 
each computational cycle. No relevant change in the results was observed, 
either in the peak intensity or in its position in frequency. 
Therefore we 
conclude that the peak is associated with physical electron cyclotron modes. 

It is worth noticing how the aforementioned presence of wave modes at different 
time scales in not in contrast with a picture of fully developed turbulence. 
Previous MHD studies have proven that Eulerian frequency spectra can present 
intensity peaks at fixed frequencies associated to wave modes even when no  
signature of such modes is present in the wavenumber spectra 
\citep{dmitruk2004discrete,dmitruk2009waves}. A result similar to the one 
presented here was found recently in two dimensional hybrid simulations of 
kinetic plasma turbulence, where several peaks at frequency larger than the ion 
cyclotron frequencies have been found in the Eulerian spectra of the magnetic 
field \citep{parashar2010kinetic}. Here we show such a behavior in a turbulent 
environment that is self-consistently generated by magnetic reconnection. A 
detailed description of the modes that contribute to shape the frequency spectra 
along with the effects due to the lack of validity of Taylor hypothesis goes 
beyond the scope of this work and will be considered for future studies.

\section{Discussion and conclusion}

We presented a full kinetic 3D simulation of reconnection in the 
presence of a strong guide field with reduced mass ratio. The simulation 
considered periodic boundary conditions in the direction of the outflow allowing 
counter propagating
reconnection jets to collide. We showed that the collision of reconnection 
jets coming from two neighbouring X-lines drives 
turbulence in 
the magnetic island within them. Turbulence manifests in the continuous 
disruption and formation of new currents sheets mainly directed along the guide 
field, whose thicknesses range between the electron and ion inertial 
lengths.

The turbulence produced in the 
collision is quasi two 
dimensional, with stronger gradients in the direction pependicular 
to the guide field. Dynamical activity in the electric current density
remains well confined in a small region of 
the simulation box that occupies a small percentage (6\%) of the total volume. 
Using an idenitification method based on the the
magnetic vector potential, we numerically 
separated the turbulent region from the rest of the 
computational domain. Analysis of spacial second order structure functions 
show that the selected central region is where turbulence actually 
develops allowing an efficient cascade of magnetic energy to sub-ion scales. 
This cascade remains active for the duration of our simulation. Since 
there is no forcing except for the initial one due to jets collision, the 
current activity slightly reduces in time, remaining confined in the same 
region.

Power spectrum analysis of electric and magnetic fluctuations 
in the turbulent region reveal interesting properties that, at first sight, 
stand in contrast to a number of previous observations and numerical simulations. 
However, the assumption of Taylor hypothesis made in previous works do not 
apply to this case, which explains the apparent discrepancy. Under these conditions, the Eulerian frequency spectra computed in our analysis are quite independent of the spectra computed in the Taylor hypothesis scenario, which are interpreted as wavenumber spectra. 
There is no simple relationship between spatial structure 
and temporal structure in the present case. 
Interestingly, the analysis of Eulerian frequency spectra in the turbulent 
region reveals the presence of wave activity at different frequencies: parallel 
propagating whistler between the ion and electron cyclotron frequency, and 
electromagnetic and electrostatic modes at the electron cyclotron frequencies. 
The presence of whistlers in the region compressed by counter streaming 
reconnection jets was recently found in in-situ observation in the case of 
small guide field \citep{alexandrova2016two}. Electrostatic and electromagnetic 
wave activity at the electron cyclotron frequency has been also detected in 
previous observations of magnetic reconnection in the vicinity of the 
reconnection sites 
\citep{tang2013themis,viberg2013mapping,khotyaintsev2016electron}. 
Here, we observed electron cyclotron wave activity in the turbulent region 
formed by reconnection jets collision.

In a recent paper, \citet{oieroset2016mms} have shown the first observation of 
reconnection jets collision in the presence of a strong guide field. The 
dynamical picture presented in that work imagined an event of secondary 
reconnection activated by the counter propagating jets both carrying magnetic 
field lines with opposite polarities. Our simulation shows that the physical 
situation can be less laminar than predicted and that turbulence develops after 
jets collision producing several current sheets, secondary reconnection 
events, and wave modes up to the electron cyclotron frequency.
Our results strongly agrees with the recent results 
by \citet{fu2017intermittent}, which have found turbulent magnetic reconnection 
in the region surrounding an O-point.
The present results are likely to be 
relevant for the interpretation of future 
observations of collisions of reconnection jets, which may have distinctive signatures in space and astrophysical plasmas.

\acknowledgments

The present work was supported by the H2020 Project DEEP-ER and DEEP-EST, by the Onderzoekfonds KU Leuven (Research Fund KU Leuven, GOA scheme and Space Weaves RUN project). This research used resources of the National Energy Research Scientific Computing Center, which is supported by the Office of Science of the US Department of Energy under Contract no. DE-AC02-05CH11231. Additional computing has been
provided by NASA NAS and NCCS High Performance Computing, by the Flemish
Supercomputing Center (VSC) and by PRACE Tier-0 allocations.

\appendix

\section{Appendix information}





\end{document}